\pdfoutput=1
\documentclass[11pt]{article}

\usepackage[preprint]{acl}

\usepackage{times}
\usepackage{latexsym}
\usepackage[T1]{fontenc}
\usepackage[utf8]{inputenc}
\usepackage{microtype}
\usepackage{inconsolata}
\usepackage{graphicx}
\usepackage{booktabs}

\title{What Do Audio-Language Judges of Generated Music Listen For?\\Four Probes with Controls from Expert Ratings}

\author{Mateusz Modrzejewski \quad Tomasz Radzikowski \\
        Institute of Computer Science, Warsaw University of Technology \\
        \texttt{\{mateusz.modrzejewski, tomasz.radzikowski\}@pw.edu.pl}}

\begin{document}
\maketitle

\begin{abstract}
Open audio-language models are used as judges of generated music: given a clip and its text prompt, the model
rates the clip, and the judge is validated by how well its ratings correlate with expert ratings. The
correlation does not show what the judge responds to: the music, the text prompt, or the way the question is asked.
We propose four probes that test each of these, three of them read against a control computed from the human ratings:
removing the audio, replacing the clip with another model's clip for the same prompt, swapping the order of two
clips in pairwise judging, and rewording the instruction. We apply them to five open judges on all 2,748 clips of MusicEval, and the last two to the SongEval songs. Without audio, no judge exceeds what the prompt alone predicts from the human ratings, and for prompt alignment
every judge falls below it. With audio, every judge follows the clip it hears when rating prompt alignment, two do
when rating musical quality, and the strongest judge's quality score tracks the experts' alignment rating more than
their quality rating. The pairwise verdicts of the four smaller judges follow the position of a clip in the pair (its slot) rather than
its quality on both rating sets. For musical quality, rewording the instruction moves the correlation of most judges as
much as changing the judge. MusicEval's prompt assignment bounds any prompt-only judge at the model level. We recommend
reporting the four probes with every music judge.
\end{abstract}

\section{Introduction}
\label{sec:intro}

Automatic judges now rank text-to-music models. Among them are audio-language models prompted to rate a
generated clip given its text prompt, with reported correlations with expert ratings on MusicEval
\citep{liu2025musiceval} from near zero to moderate \citep{ma2026cmi,zhang2026jastin}. A correlation measures how much a judge agrees with experts. It does not say whether the agreement comes from the
music, the text prompt, or the protocol. For judges of speech,
\citet{park2026auditing} show that shortcuts depend on the protocol and recommend a matched probe for each model
and protocol, \citet{manakul2025audiojudge} report position bias in pairwise judging, and \citet{foo2026all} find
that audio-language models retain 60--72\% of their accuracy on question-answering benchmarks without audio.

Generated music allows two probes that the speech audits did not have: experts rate a clip for musical quality
and for alignment with the prompt, and a judge can be asked for either; and many models render the same prompt, so
one model's clip can be replaced by another's while everything else stays fixed.

This paper makes two contributions. First, we define four probes for music judges, each with a control (a prompt-only
floor computed from the human ratings, a same-prompt swap, an order swap binned by the expert quality gap, and
a paraphrase range), so that a reported correlation can be checked against each source it may come from. The floor also shows that 25 of
the 31 models in MusicEval share one prompt set, which bounds any prompt-only judge at the model level. Second, we apply them to five open judges on all MusicEval clips. Every judge follows the clip it hears on prompt alignment and two do on musical quality; the pairwise verdicts of the four smaller judges follow the slot on both rating sets, and rewording the instruction
moves a quality correlation as much as changing the judge. 

\begin{figure}[t]
\centering
\includegraphics[width=\columnwidth]{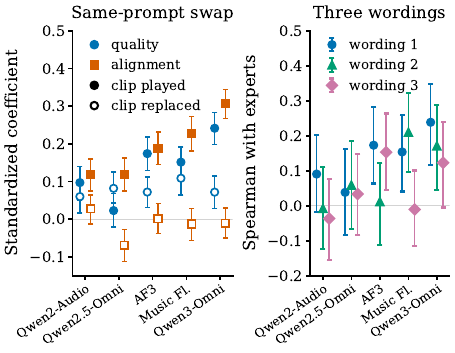}
\caption{Two of the four probes for the five judges, with 95\% bootstrap intervals. Left: in the same-prompt swap,
the standardized coefficient of the judge's score on the expert rating of the clip played (filled) and of the clip
it replaced (open), for the quality and the alignment target; a judge that listens follows the clip played. Right:
correlation with expert quality ratings under three wordings of the instruction on the same 300 clips.}
\label{fig:probes}
\end{figure}

\section{Related Work}
\label{sec:related}

\paragraph{Judges of generated music.}
Automatic scores of generated music include a distribution metric selected for its agreement with human
preferences \citep{huang2025aligning} and predictors trained on human ratings \citep{kim2026tunejury,zhu2026muq},
and one such predictor follows genre rather than quality \citep{zhang2026genre}. Zero-shot audio-language judges are
evaluated on MusicEval by \citet{ma2026cmi} and \citet{zhang2026jastin}, who report agreement without probes.

\paragraph{Shortcuts in model judges.}
Text judges are validated by agreement with human preferences \citep{zheng2023judging}, read through score-token
probabilities \citep{liu2023geval}, and move with prompt format \citep{sclar2023quantifying} and candidate
position \citep{wang2023fair}. For audio judges of speech,
\citet{manakul2025audiojudge} find that position bias grows as the quality gap narrows, and
\citet{park2026auditing} find a judge that picks the same slot regardless of order. Content-blind probes from generator identity and description length match trained judges of emotion descriptions
\citep{yang2026style}, and text-only models score well on music question answering \citep{zang2025you}. We transfer the no-audio and order probes to music and add the same-prompt swap; to our knowledge, no audit of
this kind has been published for judges of generated music.

\section{Setup}
\label{sec:setup}

\paragraph{Data.}
MusicEval \citep{liu2025musiceval} contains 2,748 clips generated by 31 text-to-music models, rated by
14 music experts on a five-point scale for overall musical impression and for alignment with the prompt.
The release has 260 distinct prompt texts (Appendix~\ref{app:instructions}): 25 models share one
set of 100 prompts, and six use prompts from their own demonstration pages. Clips longer than 30~s (48 clips of one model) are cut to their first 30~s for
every judge. A single expert agrees with the mean of the others at Spearman 0.605 over all clips and 0.252 within a
model; these are the reference levels for a judge's correlations.

SongEval \citep{yao2025songeval} provides a second rating set: 2,399 generated songs rated on five aesthetic
dimensions, released without prompts or model identifiers, so that only the order swap and the paraphrase range apply, besides
the plain audio condition. We use 30~s from the 30~s mark of each song and the mean over dimensions and raters.

\paragraph{Judges and readout.}
We evaluate Qwen2-Audio-7B-Instruct, Qwen2.5-Omni-7B, Audio Flamingo~3, Music Flamingo, and Qwen3-Omni-30B-A3B
with FP8 weights (versions in Appendix~\ref{app:models}, instructions in Appendix~\ref{app:instructions}). Each judge rates a clip from 1 to 5, and the
score is the expected value over the next-token probabilities of the five digits. On the 413 test clips this readout matches the published correlations of \citet{ma2026cmi} within 0.05 (0.198 against 0.220 for Qwen2-Audio, 0.185 against 0.145 for Qwen2.5-Omni, 0.286 against 0.310 for
Qwen3-Omni); the greedy digit gives 0.005 for Qwen2-Audio. Repeated runs give identical scores, and a second GPU type and torch build reproduce them at Spearman 0.99 or higher. The unquantized Qwen3-Omni changes no cell of Tables~\ref{tab:main} and~\ref{tab:probes} by
more than 0.02 but picks the same slot in more SongEval pairs (Table~\ref{tab:quant}).

\paragraph{Probes.}
(1) \emph{No audio}: the judge receives the instruction used with audio, with the audio omitted; a variant
replaces the audio by 30~s of digital silence \citep{foo2026all}. The control is the prompt-only floor: the rank correlation with the expert rating that the prompt's effect on human
ratings reaches, estimated from the other models' clips.
(2) \emph{Same-prompt swap}: the judge hears another model's clip for the same prompt (duration within 2~s). We
regress its standardized score on the standardized expert ratings of the clip played and of the clip replaced; a
judge that listens follows the clip played. (3) \emph{Order swap}: 600 pairs of clips for the same prompt, cropped to 12~s and concatenated
with a short tone, judged in both orders for overall quality, and compared with the judge's own pointwise ratings
of the same 12~s crops. (4) \emph{Paraphrase range}: three wordings of the instruction on 300 clips. Two protocol variants accompany it
(Appendix~\ref{app:instructions}): three answer formats (scales from 1 to 7 and from 1 to 9, and a yes/no question read as
the probability of yes) and a two-turn conversation that asks for a probability before the rating. Probes (1), (2), and (4)
run for both rating targets.

\paragraph{Statistics.}
Spearman correlations carry 95\% bootstrap intervals (1,000 resamples) over clips; model-level intervals resample
models and then clips within each model. Within-model values are means of per-model correlations over the 25 shared-prompt models. Differences are paired
bootstraps over the same clips or pairs (10,000 resamples) with Holm correction.

\section{Results}
\label{sec:results}

\begin{table}[t]
\centering
\small
\setlength{\tabcolsep}{3pt}
\begin{tabular}{@{}lccccc@{}}
\toprule
 & \multicolumn{3}{c}{Overall quality} & Align. & Model \\
\cmidrule(lr){2-4}
Judge & Audio & No audio & $\Delta_w$ & Audio & level \\
\midrule
Qwen2-Audio & 0.13 & 0.03 & 0.06 & 0.17 & 0.42 \\
Qwen2.5-Omni & 0.11 & 0.01 & 0.05 & 0.19 & 0.32 \\
Audio Flamingo 3 & 0.24 & 0.02 & 0.12 & 0.23 & 0.73 \\
Music Flamingo & 0.21 & 0.05 & 0.07 & 0.23 & 0.55 \\
Qwen3-Omni (FP8) & 0.31 & 0.04 & 0.23 & 0.34 & 0.36 \\
\midrule
Prompt-only floor & -- & 0.08 & -- & 0.28 & $\le$0.28 \\
One expert & 0.61 & -- & -- & -- & 0.98 \\
One expert, within & 0.25 & -- & -- & 0.33 & -- \\
\bottomrule
\end{tabular}
\caption{Spearman correlation with expert ratings. The Audio and No audio columns are computed over all 2,748
clips; No audio omits the audio and keeps the instruction used with audio. $\Delta_w$ is audio minus no audio within a model, and every interval is above zero. Align.\ is the
prompt-alignment target with audio, within a model. Model level is over 31 models after adjusting for each model's prompts; the expert's 0.98 is unadjusted.
The expert rows give one expert against the mean of the others, pooled or within a model. Intervals are in Table~\ref{tab:ci-main}.}
\label{tab:main}
\end{table}

\paragraph{Without audio, no more than the prompt allows.}
Removing the audio lowers the correlation with expert quality to 0.01--0.05 over all clips when the
instruction is otherwise unchanged, and to 0.00--0.04 with silence instead of audio (Tables~\ref{tab:main} and~\ref{tab:ci-main}),
compared with the 0.08 that the prompt alone reaches from human ratings. On the same clips, a paired test with
Holm correction places no judge below that floor for quality and every judge below it for alignment (0.02 to 0.07
against 0.19; Table~\ref{tab:floor-paired}). Within a model, audio adds a measurable amount for every judge and both targets under the main wording
(mean over the 25 shared-prompt models, 0.05--0.23 for quality and 0.08--0.31 for alignment). Under the other two
wordings, on 300 clips, every judge has at least one wording in which the audio gain on quality has an interval
that includes zero (Table~\ref{tab:noaudio}). Under the main wording every judge uses the audio, and without it no judge draws more from the prompt than the
human ratings allow. At the model level, a prompt-only judge would tie the 25 shared-prompt models; its
correlation over 31 models is at most 0.28.

\begin{table}[t]
\centering
\small
\setlength{\tabcolsep}{3pt}
\setlength{\tabcolsep}{1.6pt}
\begin{tabular}{@{}lcccccc@{}}
\toprule
 & \multicolumn{2}{c}{Swap, qual.} & Swap, al. & \multicolumn{3}{c}{Order swap} \\
\cmidrule(lr){2-3}\cmidrule(lr){4-4}\cmidrule(lr){5-7}
Judge & played & repl. & played & same & 2nd & agree \\
\midrule
Qwen2-Audio & 0.10 & 0.06 & 0.12 & 0.95 & 0.86 & 0.46 \\
Qwen2.5-Omni & 0.02 & 0.08 & 0.12 & 0.69 & 0.42 & 0.55 \\
Audio Flamingo 3 & 0.17 & 0.07 & 0.19 & 0.78 & 0.40 & 0.50 \\
Music Flamingo & 0.15 & 0.11 & 0.23 & 0.99 & 0.99 & 0.66 \\
Qwen3-Omni (FP8) & 0.24 & 0.07 & 0.31 & 0.25 & 0.21 & 0.67 \\
\bottomrule
\end{tabular}
\caption{The same-prompt swap columns give standardized regression coefficients of the judge's score on the expert
rating of the clip played and of the clip it replaced (2,327 swaps). The order-swap columns give the same-slot rate (the share of 600
pairs in which the judge picks the same slot in both orders, so that its preferred clip changes with the order), the
share in which that slot is the second, and the share in which the order-averaged preference agrees with the experts
(chance: 0.50, 0.25, and 0.50).
Intervals are in Tables~\ref{tab:ci-probes} and~\ref{tab:ci-order}.}
\label{tab:probes}
\end{table}

\paragraph{The audio is followed for alignment, less for quality.}
In the same-prompt swap (Figure~\ref{fig:probes}, Table~\ref{tab:probes}), all five judges follow the clip they
hear when they rate
alignment, and the rating of the clip they replaced has no positive effect. A paired bootstrap of the difference between the two coefficients, with Holm correction over the five judges,
confirms this for every judge on alignment. When they rate quality, the difference is above zero for Audio
Flamingo~3 (0.10) and Qwen3-Omni (0.17) only; Qwen2.5-Omni's quality score does not follow the clip it hears
(difference $-0.06$, interval including zero) (Table~\ref{tab:swap-paired}). Playing a clip under the wrong prompt lowers the alignment scores of the four smaller judges by up to 0.46
points and their quality scores by at most 0.11, and it lowers both scores of Qwen3-Omni, by 1.57 and 1.05 points
(Table~\ref{tab:ci-probes}). A within-model decomposition agrees: with both expert ratings as predictors, Qwen2.5-Omni's quality score loads
on expert alignment (0.17) and negatively on expert quality ($-0.07$), the Flamingo judges load more on quality
than on alignment, Qwen2-Audio loads on both about equally, and Qwen3-Omni's quality score loads on expert alignment (0.37)
more than on expert quality (0.07) (Table~\ref{tab:ci-probes}; the two expert ratings correlate at 0.74 within a
model).

\paragraph{Pairwise verdicts of the smaller judges follow the slot.}
With the two clips concatenated, the four smaller judges pick the same slot in both orders in 69--99\% of pairs
(Table~\ref{tab:probes}); Music Flamingo picks the second clip in 99\% of pairs. Qwen3-Omni does so in 25\% of pairs and agrees with the experts in 67\%. For the four smaller judges the same-slot rate is nearly constant across the expert quality gap: binned by the
gap, each judge's rates differ by at most 0.05. Qwen3-Omni shows the pattern reported for speech judges: its rate falls
from 0.30 for small gaps to 0.19 for the largest (Table~\ref{tab:ci-order}), and its Thinking variant picks the
same slot in 35\% to 41\% of pairs in every gap bin. For the two smaller judges tested with the two clips as separate inputs, the slot preference does not come from
concatenation: Qwen2.5-Omni's same-slot rate rises from 0.69 to 0.77, and Qwen2-Audio picks the same slot in 0.97
of pairs and the second clip in 0.94; Qwen3-Omni's rate falls from 0.25 to 0.19. On 300 SongEval pairs, the four smaller
judges pick the same slot in 0.59 to 1.00 of pairs and Qwen3-Omni in 0.54. Rated one clip at a time, Audio Flamingo~3 and
Qwen2-Audio order the same pairs as the experts do in 0.66 and 0.58 of cases, against 0.50 and 0.46 for their
pairwise verdicts; Qwen3-Omni's pairwise verdict is the better of its two by 0.05 (Table~\ref{tab:consistency}).

\paragraph{For quality, the wording moves the result as much as the judge.}
Across three wordings of the instruction, Music Flamingo's quality correlation on 300 clips ranges from
$-0.01$ to 0.21, Audio Flamingo~3's from 0.01 to 0.17, and Qwen3-Omni's from 0.12 to 0.24 (Figure~\ref{fig:probes},
Table~\ref{tab:paraphrase}); on all 2,748 clips, Music Flamingo's ranges from 0.00 to 0.25 (paired difference 0.24
[0.21, 0.27]) and Qwen3-Omni's from 0.13 to 0.31 (Table~\ref{tab:paraphrase-full}). For four of the five judges these ranges are not distinguishable from the spread of the five judges'
correlations on the same clips, and Qwen2.5-Omni's range is smaller (Table~\ref{tab:paired-judges}). The
response format matters less among graded scales than between a scale and a yes/no question: scores on 1--5, 1--7,
and 1--9 scales correlate at 0.95 or more for four judges, while the yes/no probability correlates with them at 0.58
to 0.96, and under it Audio Flamingo~3, Music Flamingo, and Qwen3-Omni do not differ in their correlation with the experts
(paired intervals include zero; Table~\ref{tab:format}). For prompt alignment the ranges are narrower, and Qwen3-Omni's
correlation does not depend on the wording (0.34 to 0.35 on all clips).

\paragraph{In a conversation, the rating follows the judge's earlier answer.}
Asked first for the probability that the clip matches the prompt and then for the rating, three of the four smaller
judges follow their written probability more than their single-turn rating (by 0.23 to 0.37), with no gain in
agreement with the experts. The written probabilities take few distinct values: the three most frequent cover 54\% to
98\% of answers (Table~\ref{tab:chain}). Text models tuned with human feedback give better-calibrated verbal than token
probabilities \citep{tian2023just}, but here the verbal probability agrees with the experts less than the token
probability of yes for three of the four smaller judges (by 0.14 to 0.20).

\paragraph{Judge rankings do not transfer between rating sets.}
On SongEval the judges agree with the expert mean at 0.27 to 0.67, higher than on MusicEval; Qwen3-Omni has the highest correlation on both sets under the main wording (0.67 on SongEval; 0.55 to 0.67
across wordings). The two rating sets differ in the spread of the
expert ratings (standard deviation 0.89 on MusicEval against 1.12 on SongEval)
and in the instruction, which carries no prompt on SongEval. The order of the four smaller judges changes: Qwen2.5-Omni, the weakest of them on MusicEval quality, is the
strongest on SongEval (0.57 to 0.60 across wordings),
and Audio Flamingo~3's correlation ranges from 0.32 to 0.50 across wordings. A judge's agreement with experts on one rating set, and its rank among the judges, cannot be assumed to carry over
to another.

\section{Conclusion}
\label{sec:conclusion}

We propose four probes for audio-language judges of generated music, each with a control, and apply them to five
open judges. The judges respond to the audio, more reliably for prompt
alignment than for musical quality: all five follow the clip they hear when rating alignment, and two when rating
quality. The pairwise verdicts of the four smaller judges follow the slot. For musical quality, the wording of the instruction matters as much as the choice
of judge for four of the five. MusicEval's prompt assignment bounds any prompt-only judge at the model level. We recommend reporting these four probes with every music judge, as in Table~\ref{tab:metric-card}.

\section*{Limitations}

We probe open judges that fit one workstation, and closed judges may behave differently. The largest judge runs
with FP8 weights from a community quantization (Section~\ref{sec:setup}). The main results come from MusicEval, whose
clips are at most 30~s and whose prompts may be part of the judges' training data. SongEval supports only two of the
four probes, and we use a 30~s excerpt of each song. Expert quality and alignment ratings are
correlated, which limits the decomposition. Each clip is judged once per condition. The order swap uses clips cropped to 12~s and asks only for overall
quality, and separate inputs cover three of the five judges. The main tables use one wording of the instruction, and
the paraphrase range for three judges, the answer formats, and the conversation probe rest on 300 clips. Model-level correlations rest on 31
models and have wide intervals.

\section*{Ethical Considerations}

We use MusicEval (CC BY-NC 4.0) and SongEval (CC BY-NC-SA 4.0) under their licenses for research only; no new human ratings were collected, and we do
not redistribute audio. Automatic judges that score the prompt rather than the music could steer the development of text-to-music models toward prompts that judges favor; the probes are meant to detect this. The judges'
licenses (Apache-2.0 and NVIDIA non-commercial) permit this research use. We used an AI assistant to help write and debug code, search the literature, and edit
the text; the authors designed the study, checked the results and citations, and take responsibility for the content.

\bibliography{references}

\appendix
\setcounter{dbltopnumber}{6}\renewcommand{\dbltopfraction}{0.95}\renewcommand{\dblfloatpagefraction}{0.9}
\renewcommand{\topfraction}{0.95}\renewcommand{\textfraction}{0.05}

\section{Judges, Versions, and Decoding}
\label{app:models}

\begin{table*}[tp]
\centering
\footnotesize
\begin{tabular}{@{}lp{4.6cm}llll@{}}
\toprule
Judge & Hugging Face identifier & Revision & Params & Precision & Software \\
\midrule
Qwen2-Audio & \texttt{Qwen/Qwen2-Audio-7B-Instruct} & \texttt{0a095220c3} & 8.4B & bf16 & transf.\ 5.17.0 \\
Qwen2.5-Omni & \texttt{Qwen/Qwen2.5-Omni-7B} & \texttt{ae9e169054} & 10.7B & bf16 & transf.\ 5.17.0 \\
Audio Flamingo 3 & \texttt{nvidia/audio-flamingo-3-hf} & \texttt{7d4bae64ee} & 8.3B & bf16 & transf.\ 5.17.0 \\
Music Flamingo & \texttt{nvidia/music-flamingo-2601-hf} & \texttt{6b5be086d5} & 8.3B & bf16 & transf.\ 5.17.0 \\
Qwen3-Omni & \texttt{sammysun0711/Qwen3-Omni-30B-}\newline\texttt{A3B-Instruct-FP8-Dynamic} & \texttt{622f7242c4} & 35.3B (MoE) & FP8 (dynamic) & vLLM 0.30 \\
\bottomrule
\end{tabular}
\caption{Judges with their Hugging Face revision (first ten characters), parameters, precision, and software (transf.: transformers). The score is the expected value over the digits 1 to 5 (the letters A and B in pairwise judging) under greedy decoding. Qwen3-Omni is a community FP8 quantization of \texttt{Qwen/Qwen3-Omni-30B-A3B-Instruct} on two RTX 4090 GPUs; the other judges run on one.}
\label{tab:models}
\end{table*}

\section{Intervals and Further Tables}
\label{app:tables}

Intervals for Tables~\ref{tab:main} and~\ref{tab:probes}, and further probes. Every interval is a 95\% bootstrap
interval over clips, pairs, or models as the table states; paired differences use 10,000 resamples and Holm
correction within the family named in the caption.

\begin{table*}[tp]
\centering
\scriptsize
\begin{minipage}[t]{0.59\textwidth}\centering\setlength{\tabcolsep}{2pt}
(a)\\[2pt]
\begin{tabular}{@{}llccc@{}}
\toprule
Comparison & Target & Correlations & Difference & $p_{\mathrm{Holm}}$ \\
\midrule
\multicolumn{5}{@{}l}{\emph{MusicEval, all clips}} \\
Qwen2-Audio -- Qwen2.5-Omni & quality & 0.13 -- 0.11 & 0.02 [-0.02, 0.05] & 0.362 \\
Qwen2-Audio -- Audio Flamingo 3 & quality & 0.13 -- 0.24 & -0.11 [-0.15, -0.08] & $<$0.001 \\
Qwen2-Audio -- Music Flamingo & quality & 0.13 -- 0.21 & -0.09 [-0.12, -0.05] & $<$0.001 \\
Qwen2-Audio -- Qwen3-Omni (FP8) & quality & 0.13 -- 0.31 & -0.18 [-0.22, -0.13] & $<$0.001 \\
Qwen2.5-Omni -- Audio Flamingo 3 & quality & 0.11 -- 0.24 & -0.13 [-0.16, -0.10] & $<$0.001 \\
Qwen2.5-Omni -- Music Flamingo & quality & 0.11 -- 0.21 & -0.10 [-0.13, -0.07] & $<$0.001 \\
Qwen2.5-Omni -- Qwen3-Omni (FP8) & quality & 0.11 -- 0.31 & -0.19 [-0.23, -0.15] & $<$0.001 \\
Audio Flamingo 3 -- Music Flamingo & quality & 0.24 -- 0.21 & 0.03 [0.00, 0.06] & 0.204 \\
Audio Flamingo 3 -- Qwen3-Omni (FP8) & quality & 0.24 -- 0.31 & -0.07 [-0.11, -0.03] & 0.012 \\
Music Flamingo -- Qwen3-Omni (FP8) & quality & 0.21 -- 0.31 & -0.09 [-0.13, -0.05] & $<$0.001 \\
Qwen2-Audio -- Qwen2.5-Omni & alignment & 0.13 -- 0.10 & 0.03 [-0.01, 0.07] & 0.289 \\
Qwen2-Audio -- Audio Flamingo 3 & alignment & 0.13 -- 0.16 & -0.03 [-0.07, 0.01] & 0.289 \\
Qwen2-Audio -- Music Flamingo & alignment & 0.13 -- 0.20 & -0.07 [-0.11, -0.03] & 0.005 \\
Qwen2-Audio -- Qwen3-Omni (FP8) & alignment & 0.13 -- 0.35 & -0.22 [-0.26, -0.18] & $<$0.001 \\
Qwen2.5-Omni -- Audio Flamingo 3 & alignment & 0.10 -- 0.16 & -0.06 [-0.09, -0.03] & 0.007 \\
Qwen2.5-Omni -- Music Flamingo & alignment & 0.10 -- 0.20 & -0.10 [-0.14, -0.06] & $<$0.001 \\
Qwen2.5-Omni -- Qwen3-Omni (FP8) & alignment & 0.10 -- 0.35 & -0.25 [-0.28, -0.21] & $<$0.001 \\
Audio Flamingo 3 -- Music Flamingo & alignment & 0.16 -- 0.20 & -0.04 [-0.07, -0.01] & 0.061 \\
Audio Flamingo 3 -- Qwen3-Omni (FP8) & alignment & 0.16 -- 0.35 & -0.19 [-0.22, -0.15] & $<$0.001 \\
Music Flamingo -- Qwen3-Omni (FP8) & alignment & 0.20 -- 0.35 & -0.15 [-0.18, -0.11] & $<$0.001 \\
\multicolumn{5}{@{}l}{\emph{Wording range minus the spread of the five judges, 300 clips}} \\
Qwen2-Audio & quality & 0.13 -- 0.20 & -0.07 [-0.23, 0.06] & 0.793 \\
Qwen2.5-Omni & quality & 0.03 -- 0.20 & -0.18 [-0.30, -0.06] & 0.022 \\
Audio Flamingo 3 & quality & 0.16 -- 0.20 & -0.04 [-0.19, 0.08] & 1.000 \\
Music Flamingo & quality & 0.22 -- 0.20 & 0.02 [-0.14, 0.14] & 1.000 \\
Qwen3-Omni (FP8) & quality & 0.12 -- 0.20 & -0.08 [-0.22, 0.02] & 0.508 \\
Qwen2-Audio & alignment & 0.06 -- 0.28 & -0.22 [-0.36, -0.06] & 0.029 \\
Qwen2.5-Omni & alignment & 0.19 -- 0.28 & -0.09 [-0.22, 0.04] & 0.602 \\
Audio Flamingo 3 & alignment & 0.02 -- 0.28 & -0.26 [-0.36, -0.11] & 0.002 \\
Music Flamingo & alignment & 0.09 -- 0.28 & -0.19 [-0.31, -0.06] & 0.016 \\
Qwen3-Omni (FP8) & alignment & 0.01 -- 0.28 & -0.27 [-0.39, -0.14] & $<$0.001 \\
\bottomrule
\end{tabular}
\end{minipage}\hfill
\begin{minipage}[t]{0.405\textwidth}\centering\setlength{\tabcolsep}{1.2pt}
(b)\\[2pt]
\begin{tabular}{@{}llccc@{}}
\toprule
Judge & Target & No audio & Difference & $p_{\mathrm{Holm}}$ \\
\midrule
Qwen2-Audio & qual. & 0.07 & -0.01 [-0.06, 0.04] & 1.000 \\
Qwen2.5-Omni & qual. & 0.01 & -0.06 [-0.12, -0.01] & 0.107 \\
Audio Flamingo 3 & qual. & 0.04 & -0.04 [-0.09, 0.01] & 0.555 \\
Music Flamingo & qual. & 0.06 & -0.02 [-0.07, 0.03] & 1.000 \\
Qwen3-Omni (FP8) & qual. & 0.05 & -0.03 [-0.08, 0.02] & 0.710 \\
Qwen2-Audio & align. & 0.07 & -0.11 [-0.16, -0.07] & $<$0.001 \\
Qwen2.5-Omni & align. & 0.05 & -0.13 [-0.18, -0.08] & $<$0.001 \\
Audio Flamingo 3 & align. & 0.06 & -0.13 [-0.18, -0.08] & $<$0.001 \\
Music Flamingo & align. & 0.05 & -0.14 [-0.19, -0.09] & $<$0.001 \\
Qwen3-Omni (FP8) & align. & 0.02 & -0.16 [-0.22, -0.11] & $<$0.001 \\
\bottomrule
\end{tabular}\\[8pt]
(c)\\[2pt]
\begin{tabular}{@{}llcc@{}}
\toprule
Judge & Target & played $-$ replaced & $p_{\mathrm{Holm}}$ \\
\midrule
Qwen2-Audio & qual. & 0.04 [-0.04, 0.11] & 0.497 \\
Qwen2.5-Omni & qual. & -0.06 [-0.14, 0.02] & 0.373 \\
Audio Flamingo 3 & qual. & 0.10 [0.03, 0.17] & 0.027 \\
Music Flamingo & qual. & 0.04 [-0.03, 0.11] & 0.497 \\
Qwen3-Omni (FP8) & qual. & 0.17 [0.09, 0.24] & 0.000 \\
Qwen2-Audio & align. & 0.09 [0.02, 0.16] & 0.008 \\
Qwen2.5-Omni & align. & 0.19 [0.12, 0.26] & 0.000 \\
Audio Flamingo 3 & align. & 0.19 [0.12, 0.25] & 0.000 \\
Music Flamingo & align. & 0.24 [0.17, 0.31] & 0.000 \\
Qwen3-Omni (FP8) & align. & 0.32 [0.25, 0.38] & 0.000 \\
\bottomrule
\end{tabular}
\end{minipage}
\caption{Paired tests, with Holm-corrected $p$ and 95\% paired bootstrap intervals. (a) The judge and wording comparisons in the text: differences of Spearman correlations on the same clips, adjusted within each block and target. (b) No-audio correlation (instruction unchanged) against the prompt-only floor (0.08 for quality, 0.19 for alignment) on the 2,501 clips where the floor is defined, adjusted over the ten tests. (c) Same-prompt swap: coefficient of the clip played minus that of the clip replaced (2,327 swaps), adjusted over the five judges of each target.}
\label{tab:paired-judges}\label{tab:floor-paired}\label{tab:swap-paired}
\end{table*}

\begin{table*}[tp]
\centering
\scriptsize
\setlength{\tabcolsep}{2.5pt}
\begin{tabular}{@{}lccccc@{}}
\toprule
 & FP8 & \multicolumn{2}{c}{Unquantized (bf16)} & \multicolumn{2}{c}{Thinking} \\
\cmidrule(lr){3-4}\cmidrule(lr){5-6}
Cell & value & value & difference & value & difference \\
\midrule
pooled, audio, overall & 0.31 & 0.31 & 0.00 [0.00, 0.01] & 0.38 & 0.07 [0.04, 0.10] \\
pooled, no audio (instruction unchanged), overall & 0.04 & 0.04 & 0.01 [0.00, 0.02] & -- & -- \\
within model, audio, overall & 0.30 & 0.30 & 0.00 [0.00, 0.01] & 0.26 & -0.04 [-0.07, -0.01] \\
within-model audio gain ($\Delta_w$), overall & 0.23 & 0.22 & -0.01 [-0.02, 0.00] & -- & -- \\
swap, clip played, overall & 0.24 & 0.25 & 0.01 [0.00, 0.01] & 0.32 & 0.08 [0.04, 0.12] \\
swap, clip replaced, overall & 0.07 & 0.07 & 0.00 [0.00, 0.01] & 0.10 & 0.02 [-0.02, 0.06] \\
pooled, audio, alignment & 0.35 & 0.36 & 0.01 [0.00, 0.01] & 0.29 & -0.06 [-0.08, -0.03] \\
pooled, no audio (instruction unchanged), alignment & 0.01 & 0.01 & 0.00 [0.00, 0.01] & -- & -- \\
within model, audio, alignment & 0.34 & 0.35 & 0.00 [0.00, 0.01] & 0.30 & -0.04 [-0.07, -0.01] \\
within-model audio gain ($\Delta_w$), alignment & 0.31 & 0.31 & -0.01 [-0.02, 0.01] & -- & -- \\
swap, clip played, alignment & 0.31 & 0.32 & 0.01 [0.00, 0.02] & 0.21 & -0.10 [-0.14, -0.07] \\
swap, clip replaced, alignment & -0.01 & -0.01 & 0.00 [-0.01, 0.01] & 0.01 & 0.02 [-0.02, 0.05] \\
order swap, same-slot rate, MusicEval concatenated & 0.24 & 0.23 & -0.01 [-0.04, 0.01] & 0.40 & 0.16 [0.11, 0.21] \\
order swap, second-slot rate, MusicEval concatenated & 0.17 & 0.17 & 0.01 [-0.01, 0.03] & 0.23 & 0.06 [0.02, 0.11] \\
order swap, order-averaged agreement, MusicEval concatenated & 0.67 & 0.66 & -0.01 [-0.03, 0.01] & 0.63 & -0.03 [-0.07, 0.01] \\
order swap, same-slot rate, SongEval & 0.53 & 0.61 & 0.08 [0.03, 0.13] & -- & -- \\
order swap, order-averaged agreement, SongEval & 0.67 & 0.64 & -0.03 [-0.06, 0.01] & -- & -- \\
correlation with expert mean (SongEval), first wording & 0.67 & 0.67 & 0.00 [0.00, 0.01] & -- & -- \\
correlation with expert mean (SongEval), second wording & 0.66 & 0.66 & 0.00 [0.00, 0.01] & -- & -- \\
correlation with expert mean (SongEval), third wording & 0.55 & 0.56 & 0.01 [0.00, 0.01] & -- & -- \\
\bottomrule
\end{tabular}
\caption{Qwen3-Omni-30B-A3B unquantized (bf16) and Thinking against the FP8 judge of the main tables, on identical inputs: the cells of Tables~\ref{tab:main} and~\ref{tab:probes} and the SongEval correlations, with the difference from FP8. Unquantized and FP8 clip scores correlate at 0.91 to 0.99. The Thinking readout follows its reasoning (up to 4,096 tokens); 128 of its 23,140 MusicEval judgments, mostly without audio, give no answer and are excluded.}
\label{tab:quant}
\end{table*}

\begin{table*}[tp]
\centering
\scriptsize
\setlength{\tabcolsep}{3pt}
\begin{tabular}{@{}lccccccc@{}}
\toprule
 & & \multicolumn{2}{c}{No audio (Table~\ref{tab:main})} & \multicolumn{2}{c}{No audio, explicit wording} & \multicolumn{2}{c}{Silence} \\
\cmidrule(lr){3-4}\cmidrule(lr){5-6}\cmidrule(lr){7-8}
Judge & Audio & pooled & $\Delta_w$ & pooled & $\Delta_w$ & pooled & $\Delta_w$ \\
\midrule
Qwen2-Audio & 0.13 [0.09, 0.16] & 0.03 & 0.06 [0.04, 0.08] & 0.03 & 0.10 [0.07, 0.13] & 0.04 & 0.07 [0.05, 0.10] \\
Qwen2.5-Omni & 0.11 [0.08, 0.15] & 0.01 & 0.05 [0.03, 0.08] & 0.00 & 0.05 [0.03, 0.07] & 0.00 & 0.03 [0.01, 0.05] \\
Audio Flamingo 3 & 0.24 [0.21, 0.27] & 0.02 & 0.12 [0.09, 0.15] & 0.02 & 0.10 [0.07, 0.14] & 0.01 & 0.12 [0.09, 0.16] \\
Music Flamingo & 0.21 [0.18, 0.25] & 0.05 & 0.07 [0.04, 0.10] & 0.04 & 0.07 [0.04, 0.10] & 0.02 & 0.09 [0.07, 0.11] \\
Qwen3-Omni (FP8) & 0.31 [0.27, 0.34] & 0.04 & 0.23 [0.18, 0.28] & 0.02 & 0.23 [0.18, 0.29] & 0.00 & 0.28 [0.22, 0.34] \\
\midrule
 & \multicolumn{7}{c}{Prompt alignment, within a model} \\
\midrule
Qwen2-Audio & 0.17 & 0.09 & 0.08 [0.05, 0.11] & 0.09 & 0.09 [0.05, 0.12] & -- & -- \\
Qwen2.5-Omni & 0.19 & 0.06 & 0.13 [0.10, 0.16] & 0.09 & 0.10 [0.06, 0.14] & -- & -- \\
Audio Flamingo 3 & 0.23 & 0.07 & 0.16 [0.13, 0.19] & 0.08 & 0.16 [0.11, 0.21] & -- & -- \\
Music Flamingo & 0.23 & 0.06 & 0.18 [0.14, 0.21] & 0.05 & 0.19 [0.14, 0.23] & -- & -- \\
Qwen3-Omni (FP8) & 0.34 & 0.03 & 0.31 [0.26, 0.36] & 0.06 & 0.29 [0.23, 0.34] & -- & -- \\
\midrule
 & \multicolumn{7}{c}{Model level, overall quality with audio} \\
\midrule
Qwen2-Audio & 0.42 [-0.02, 0.71] & & & & & & \\
Qwen2.5-Omni & 0.32 [-0.07, 0.64] & & & & & & \\
Audio Flamingo 3 & 0.73 [0.50, 0.85] & & & & & & \\
Music Flamingo & 0.55 [0.21, 0.78] & & & & & & \\
Qwen3-Omni (FP8) & 0.36 [-0.05, 0.75] & & & & & & \\
\bottomrule
\end{tabular}
\caption{Table~\ref{tab:main} with intervals (over models for the model level; paired for $\Delta_w$, audio minus control within a model). The controls omit the audio under the same instruction, omit it under the explicit no-audio wording (Appendix~\ref{app:instructions}), or play 30~s of silence.}
\label{tab:ci-main}
\end{table*}

\begin{table*}[tp]
\centering
\footnotesize
\setlength{\tabcolsep}{4pt}
\begin{tabular}{@{}llcccccc@{}}
\toprule
 & & \multicolumn{4}{c}{Correlation with experts, with audio} & \multicolumn{2}{c}{Agreement between formats} \\
\cmidrule(lr){3-6}\cmidrule(lr){7-8}
Judge & Target & 1--5 & 1--7 & 1--9 & yes/no & graded & yes/no vs.\ graded \\
\midrule
Qwen2-Audio & quality & 0.09 & 0.11 & 0.08 & 0.07 & 0.76--0.86 & 0.58--0.62 \\
Qwen2.5-Omni & quality & 0.04 & 0.07 & 0.04 & 0.07 & 0.98 & 0.78 \\
Audio Flamingo 3 & quality & 0.17 & 0.19 & 0.18 & 0.32 & 0.96--0.98 & 0.67--0.72 \\
Music Flamingo & quality & 0.15 & 0.18 & 0.21 & 0.26 & 0.96--0.98 & 0.81--0.82 \\
Qwen3-Omni (FP8) & quality & 0.24 & 0.25 & 0.24 & 0.23 & 0.98--0.99 & 0.95--0.96 \\
\midrule
Qwen2-Audio & alignment & 0.10 & 0.15 & 0.14 & 0.19 & 0.72--0.82 & 0.43--0.59 \\
Qwen2.5-Omni & alignment & 0.16 & 0.18 & 0.18 & 0.30 & 0.98--0.99 & 0.82--0.83 \\
Audio Flamingo 3 & alignment & 0.16 & 0.22 & 0.22 & 0.24 & 0.97--0.98 & 0.76--0.79 \\
Music Flamingo & alignment & 0.21 & 0.23 & 0.24 & 0.29 & 0.98--0.99 & 0.86--0.87 \\
Qwen3-Omni (FP8) & alignment & 0.38 & 0.38 & 0.38 & 0.33 & 0.99--1.00 & 0.93 \\
\bottomrule
\end{tabular}
\caption{Answer formats on 300 clips, with audio: scales read as expected values over the digits, yes/no as the probability of Yes over Yes and No. Last two columns: range of correlations between a judge's scores under two formats, among the graded scales and between yes/no and each scale.}
\label{tab:format}
\end{table*}

\begin{table*}[tp]
\centering
\footnotesize
\setlength{\tabcolsep}{2pt}
\begin{tabular}{@{}lcccccc@{}}
\toprule
 & \multicolumn{2}{c}{Turn 1: verbal probability} & \multicolumn{2}{c}{Turn 2: rating from 1 to 5} & \multicolumn{2}{c}{Turn 2, replaced clip} \\
\cmidrule(lr){2-3}\cmidrule(lr){4-5}\cmidrule(lr){6-7}
Judge & top 3 & vs.\ P(yes) & own prob.\ vs.\ single & vs.\ single & own prob. & clip played \\
\midrule
Qwen2-Audio & 0.54 & -0.09 [-0.21, 0.03] & 0.37 [0.28, 0.46] & -0.01 [-0.12, 0.10] & 0.27 [0.15, 0.38] & -0.03 [-0.16, 0.11] \\
Qwen2.5-Omni & 0.98 & -0.20 [-0.31, -0.09] & 0.30 [0.23, 0.39] & 0.03 [-0.07, 0.13] & 0.19 [0.04, 0.35] & 0.14 [0.01, 0.27] \\
Audio Flamingo 3 & 0.94 & -0.16 [-0.29, -0.05] & 0.04 [-0.04, 0.13] & 0.10 [0.00, 0.20] & -- & -- \\
Music Flamingo & 0.82 & -0.14 [-0.22, -0.05] & 0.23 [0.16, 0.31] & 0.03 [-0.06, 0.13] & -- & -- \\
Qwen3-Omni (FP8) & 0.77 & 0.01 [-0.04, 0.06] & -0.01 [-0.03, 0.00] & -0.02 [-0.05, 0.01] & 0.58 [0.48, 0.68] & 0.33 [0.22, 0.43] \\
\bottomrule
\end{tabular}
\caption{Probability, then rating, in one conversation (alignment, 300 clips). Turn~1 asks for the probability that the clip matches the prompt, written as a number between 0 and 1; turn~2 repeats the rating question, alone or with a clip attached. Top 3: share of the three most frequent probabilities. Vs.\ P(yes): expert correlation of the verbal probability minus that of the yes/no token probability. Own prob.\ vs.\ single: the turn-2 rating's correlation with the turn-1 probability minus that with the single-turn rating. Vs.\ single: turn-2 minus single-turn expert correlation. Replaced clip: coefficients of the turn-2 rating on the turn-1 probability and on the expert rating of another model's clip attached in turn~2.}
\label{tab:chain}
\end{table*}

\begin{table*}[tp]
\centering
\footnotesize
\setlength{\tabcolsep}{4pt}
\begin{tabular}{@{}llcccccc@{}}
\toprule
 & & \multicolumn{2}{c}{Agrees with own pointwise} & \multicolumn{4}{c}{Agrees with the experts} \\
\cmidrule(lr){3-4}\cmidrule(lr){5-8}
Judge & Input & one order & averaged & pointwise & pairwise & difference & $p_{\mathrm{Holm}}$ \\
\midrule
Qwen2-Audio & concatenated & 0.50 & 0.48 & 0.58 & 0.46 & 0.11 [0.06, 0.17] & 0.002 \\
Qwen2.5-Omni & concatenated & 0.56 & 0.61 & 0.60 & 0.55 & 0.05 [0.00, 0.10] & 0.148 \\
Audio Flamingo 3 & concatenated & 0.51 & 0.54 & 0.66 & 0.50 & 0.16 [0.11, 0.22] & 0.000 \\
Music Flamingo & concatenated & 0.50 & 0.68 & 0.63 & 0.66 & -0.03 [-0.08, 0.02] & 0.193 \\
Qwen3-Omni (FP8) & concatenated & 0.77 & 0.81 & 0.61 & 0.67 & -0.05 [-0.09, -0.02] & 0.017 \\
\midrule
Qwen2-Audio & separate & 0.50 & 0.53 & 0.58 & 0.46 & 0.12 [0.07, 0.18] & 0.000 \\
Qwen2.5-Omni & separate & 0.55 & 0.63 & 0.60 & 0.58 & 0.02 [-0.03, 0.07] & 0.375 \\
Qwen3-Omni (FP8) & separate & 0.78 & 0.81 & 0.61 & 0.65 & -0.03 [-0.07, 0.00] & 0.168 \\
\bottomrule
\end{tabular}
\caption{Pointwise against pairwise verdicts on the 600 order-swap pairs, with pointwise ratings of the same 12~s crops (0.91 to 0.96 correlation with full-clip ratings). Agreement with the experts uses pairs with a nonzero expert gap; the difference is pointwise minus pairwise, Holm-adjusted per input type.}
\label{tab:consistency}
\end{table*}

\begin{table*}[tp]
\centering
\footnotesize
\setlength{\tabcolsep}{4pt}
\begin{tabular}{@{}llccc@{}}
\toprule
Judge & Target & Wording 1 & Wording 2 & Wording 3 \\
\midrule
Qwen2-Audio & quality & 0.09 [-0.02, 0.20] & -0.01 [-0.12, 0.11] & -0.04 [-0.16, 0.08] \\
 & alignment & 0.10 [-0.02, 0.20] & 0.04 [-0.08, 0.15] & 0.06 [-0.06, 0.16] \\
Qwen2.5-Omni & quality & 0.04 [-0.08, 0.16] & 0.06 [-0.07, 0.18] & 0.03 [-0.08, 0.15] \\
 & alignment & 0.16 [0.05, 0.27] & 0.19 [0.08, 0.30] & 0.00 [-0.12, 0.12] \\
Audio Flamingo 3 & quality & 0.17 [0.06, 0.28] & 0.01 [-0.11, 0.12] & 0.15 [0.05, 0.26] \\
 & alignment & 0.16 [0.05, 0.26] & 0.14 [0.02, 0.25] & 0.16 [0.05, 0.27] \\
Music Flamingo & quality & 0.15 [0.04, 0.26] & 0.21 [0.10, 0.32] & -0.01 [-0.11, 0.10] \\
 & alignment & 0.21 [0.10, 0.32] & 0.20 [0.08, 0.31] & 0.12 [0.01, 0.24] \\
Qwen3-Omni (FP8) & quality & 0.24 [0.12, 0.35] & 0.17 [0.05, 0.29] & 0.12 [0.00, 0.24] \\
 & alignment & 0.38 [0.27, 0.47] & 0.38 [0.27, 0.48] & 0.38 [0.27, 0.48] \\
\bottomrule
\end{tabular}
\caption{Correlation with expert ratings under three wordings, with audio, on the same 300 clips.}
\label{tab:paraphrase}
\end{table*}

\begin{table*}[tp]
\centering
\footnotesize
\setlength{\tabcolsep}{3pt}
\begin{tabular}{@{}lccccc@{}}
\toprule
 & Audio gain & Swap: played & Same-slot & Wording range & Pairwise vs.\ own \\
Judge & (within) & / replaced & rate & (clips) & pointwise \\
\midrule
Qwen2-Audio & 0.06 [0.04, 0.08] & 0.10 / 0.06 & 0.95 [0.93, 0.97] & 0.13 (300) & 0.48 [0.44, 0.52] \\
Qwen2.5-Omni & 0.05 [0.03, 0.08] & 0.02 / 0.08 & 0.69 [0.65, 0.72] & 0.03 (300) & 0.61 [0.57, 0.65] \\
Audio Flamingo 3 & 0.12 [0.09, 0.15] & 0.17 / 0.07 & 0.78 [0.75, 0.81] & 0.16 (300) & 0.54 [0.50, 0.58] \\
Music Flamingo & 0.07 [0.04, 0.10] & 0.15 / 0.11 & 0.99 [0.98, 1.00] & 0.24 (all) & 0.68 [0.65, 0.72] \\
Qwen3-Omni (FP8) & 0.23 [0.18, 0.28] & 0.24 / 0.07 & 0.24 [0.21, 0.27] & 0.18 (all) & 0.81 [0.78, 0.84] \\
\bottomrule
\end{tabular}
\caption{Metric card for musical quality: the five diagnostics we recommend reporting. Audio gain: within-model correlation with minus without audio. Swap: coefficients of the clip played and replaced. Same-slot rate: concatenated order swap. Wording range: maximum minus minimum over three wordings. Last column: pairwise against own pointwise verdicts on the same 12~s crops.}
\label{tab:metric-card}
\end{table*}

\begin{table*}[tp]
\centering
\footnotesize
\setlength{\tabcolsep}{2pt}
\begin{tabular}{@{}lcccccc@{}}
\toprule
 & \multicolumn{4}{c}{Order swap, concatenated clips} & Separate & No audio \\
\cmidrule(lr){2-5}\cmidrule(lr){6-6}\cmidrule(lr){7-7}
Judge & same slot & gap $<$0.4 & 0.4--0.8 & $\ge$0.8 & same slot & + name \\
\midrule
Qwen2-Audio & 0.95 [0.93, 0.97] & 0.93 & 0.93 & 0.96 & 0.97 [0.95, 0.98] & -0.05 [-0.09, -0.02] \\
Qwen2.5-Omni & 0.69 [0.65, 0.72] & 0.66 & 0.71 & 0.69 & 0.77 [0.74, 0.81] & 0.03 [-0.01, 0.06] \\
Audio Flamingo 3 & 0.78 [0.75, 0.81] & 0.77 & 0.79 & 0.78 & -- & 0.05 [0.02, 0.09] \\
Music Flamingo & 0.99 [0.98, 1.00] & 0.99 & 1.00 & 0.98 & -- & 0.08 [0.04, 0.11] \\
Qwen3-Omni (FP8) & 0.25 [0.21, 0.28] & 0.30 & 0.31 & 0.19 & 0.19 [0.16, 0.22] & 0.10 [0.06, 0.13] \\
Qwen3-Omni Thinking & 0.40 [0.36, 0.44] & 0.40 & 0.35 & 0.41 & -- & 0.04 [0.00, 0.08] \\
\bottomrule
\end{tabular}
\caption{Order swap on 600 pairs: same-slot rate overall, by expert quality gap (bins of 164, 107, and 329 pairs), and with separate inputs (the Flamingo processors take one audio input). Last column: no-audio quality correlation with the generating model's name in the instruction.}
\label{tab:ci-order}
\end{table*}

\begin{table*}[tp]
\centering
\scriptsize
\setlength{\tabcolsep}{2.5pt}
\begin{tabular}{@{}lcccccc@{}}
\toprule
 & \multicolumn{2}{c}{Same-prompt swap, quality} & \multicolumn{2}{c}{Cross-prompt drop} & \multicolumn{2}{c}{Decomposition} \\
\cmidrule(lr){2-3}\cmidrule(lr){4-5}\cmidrule(lr){6-7}
Judge & played & replaced & quality & align. & on quality & on align. \\
\midrule
Qwen2-Audio & 0.10 [0.05, 0.14] & 0.06 [0.02, 0.11] & 0.04 [0.02, 0.06] & 0.06 [0.05, 0.08] & 0.10 [0.04, 0.15] & 0.11 [0.05, 0.16] \\
Qwen2.5-Omni & 0.02 [-0.02, 0.07] & 0.08 [0.04, 0.13] & 0.11 [0.09, 0.14] & 0.46 [0.42, 0.51] & -0.07 [-0.13, -0.01] & 0.17 [0.11, 0.23] \\
Audio Flamingo 3 & 0.17 [0.13, 0.22] & 0.07 [0.03, 0.11] & 0.09 [0.06, 0.11] & 0.31 [0.27, 0.36] & 0.11 [0.05, 0.16] & 0.06 [0.01, 0.12] \\
Music Flamingo & 0.15 [0.11, 0.19] & 0.11 [0.06, 0.15] & 0.11 [0.09, 0.14] & 0.41 [0.37, 0.45] & 0.12 [0.07, 0.17] & 0.06 [0.01, 0.11] \\
Qwen3-Omni (FP8) & 0.24 [0.20, 0.28] & 0.07 [0.03, 0.11] & 1.05 [0.98, 1.13] & 1.57 [1.47, 1.67] & 0.07 [0.02, 0.12] & 0.37 [0.32, 0.42] \\
\bottomrule
\end{tabular}
\caption{Same-prompt swap for quality (2,327 swaps), mean score drop under another prompt (points on the 1 to 5 scale, 999 clips), and within-model regression of the quality score on both expert ratings, which correlate at 0.74.}
\label{tab:ci-probes}
\end{table*}

\begin{table*}[tp]
\centering
\footnotesize
\setlength{\tabcolsep}{3pt}
\begin{tabular}{@{}llccccccc@{}}
\toprule
 & & \multicolumn{3}{c}{Pooled, all clips} & \multicolumn{3}{c}{Within model} & \\
\cmidrule(lr){3-5}\cmidrule(lr){6-8}
Judge & Target & Wording 1 & Wording 2 & Wording 3 & 1 & 2 & 3 & Kendall $\tau$ \\
\midrule
Music Flamingo & quality & 0.21 [0.18, 0.25] & 0.25 [0.21, 0.28] & 0.00 [-0.03, 0.04] & 0.18 & 0.18 & 0.08 & 0.47--0.67 \\
Music Flamingo & alignment & 0.20 [0.17, 0.24] & 0.18 [0.15, 0.22] & 0.11 [0.07, 0.15] & 0.23 & 0.19 & 0.14 & 0.58--0.64 \\
Qwen3-Omni (FP8) & quality & 0.31 [0.27, 0.34] & 0.21 [0.17, 0.25] & 0.13 [0.09, 0.17] & 0.30 & 0.28 & 0.23 & 0.70--0.77 \\
Qwen3-Omni (FP8) & alignment & 0.35 [0.32, 0.38] & 0.34 [0.30, 0.37] & 0.35 [0.31, 0.38] & 0.34 & 0.35 & 0.35 & 0.82--0.84 \\
\bottomrule
\end{tabular}
\caption{All three wordings on all 2,748 clips: pooled correlation with the experts, mean within-model correlation over the 25 shared-prompt models, and the range of Kendall $\tau$ between a judge's scores under two wordings.}
\label{tab:paraphrase-full}
\end{table*}

\begin{table*}[tp]
\centering
\scriptsize
\setlength{\tabcolsep}{2.5pt}
\begin{tabular}{@{}lcccccc@{}}
\toprule
 & \multicolumn{3}{c}{Quality} & \multicolumn{3}{c}{Alignment} \\
\cmidrule(lr){2-4}\cmidrule(lr){5-7}
Judge & wording 1 & wording 2 & wording 3 & wording 1 & wording 2 & wording 3 \\
\midrule
Qwen2-Audio & 0.11 [0.02, 0.19] & -0.01 [-0.08, 0.07] & -0.06 [-0.14, 0.01] & 0.11 [0.01, 0.21] & 0.00 [-0.09, 0.10] & 0.08 [0.00, 0.16] \\
Qwen2.5-Omni & 0.09 [0.00, 0.19] & 0.08 [0.01, 0.14] & 0.04 [-0.02, 0.11] & 0.11 [0.00, 0.22] & 0.17 [0.09, 0.26] & -0.02 [-0.11, 0.05] \\
Audio Flamingo 3 & 0.23 [0.14, 0.32] & 0.05 [-0.03, 0.12] & 0.20 [0.13, 0.29] & 0.18 [0.08, 0.29] & 0.15 [0.08, 0.23] & 0.21 [0.13, 0.30] \\
Music Flamingo & 0.15 [0.07, 0.22] & 0.20 [0.14, 0.25] & -0.01 [-0.10, 0.07] & 0.21 [0.10, 0.32] & 0.19 [0.12, 0.26] & 0.15 [0.06, 0.23] \\
Qwen3-Omni (FP8) & 0.27 [0.11, 0.42] & 0.16 [0.01, 0.32] & 0.07 [-0.09, 0.23] & 0.29 [0.13, 0.43] & 0.31 [0.16, 0.45] & 0.32 [0.17, 0.48] \\
\bottomrule
\end{tabular}
\caption{Audio gain per wording on 300 clips: pooled correlation with audio minus without audio under the same wording (paired).}
\label{tab:noaudio}
\end{table*}

\section{Instructions and Prompt Counts}
\label{app:instructions}

The rating instruction (first wording, quality target, with audio) is: ``You will hear a music clip that a text-to-music system generated from the text prompt below. Prompt: \{prompt\} Rate the overall musical quality of the clip on a scale from 1 (very poor) to 5 (excellent). Answer with a single digit.''
For the alignment target, the last sentence asks: ``Rate how well the clip matches the text prompt on a scale from 1 (does not match at all) to 5 (matches completely). Answer with a single digit.''
In the no-audio condition of the main tables, the audio is omitted and the instruction is unchanged; the silence condition plays 30~s of digital silence under the same instruction. The explicit no-audio wording (Table~\ref{tab:ci-main}) changes the lead sentence to ``A text-to-music system generated a music clip from the text prompt below.'' and ends with ``No audio is available to you. Give your best estimate of the rating from the text alone. Answer with a single digit.''
The answer formats keep the lead sentence and the prompt and replace the last sentence by the same question on a scale from 1 to 7 or from 1 to 9 (same anchors), or by ``Is the overall musical quality of the clip good? Answer with Yes or No.'' (alignment: ``Does the clip match the text prompt? Answer with Yes or No.''). The model-name probe prepends ``The clip was generated by the text-to-music system \{name\}.''
The pairwise instruction is: ``Two music clips were generated from the same text prompt by two different systems. You will hear them one after the other, separated by a short beep: clip A first, then clip B. Prompt: \{prompt\} Which clip has the better overall musical quality? Answer with a single letter, A or B.''; with separate inputs, its second sentence reads ``The first audio is clip A and the second audio is clip B.''
The second wording (quality target, with audio) is: ``The audio is a piece of music produced by an AI music generator for the request below. Request: "\{prompt\}" How good is the music overall? Give a score from 1 to 5, where 1 is very poor and 5 is excellent. Reply with one digit only.'' The third is: ``Act as an expert music critic and listen to the generated music. Its text prompt was: \{prompt\} Score its overall musical impression from 1 (worst) to 5 (best). Output only the score.'' For the alignment target, their last sentences ask: ``How well does the music follow the request? Give a score from 1 to 5, where 1 is not at all and 5 is completely. Reply with one digit only.'' and ``Score how well the music matches the prompt from 1 (worst) to 5 (best). Output only the score.''

\paragraph{Prompt counts.} In the MusicEval release, 25 models render the same 100 prompts, and six demonstration-only models contribute 248 clips whose prompts come from their demonstration pages. These clips have 161 distinct prompt texts, one of which also occurs among the shared prompts, so the release has 260 distinct prompt texts. The dataset paper reports 384 prompts: 80 written by its authors and 20 from MusicCaps, which form the shared set, and 284 from demonstration pages, of which the release contains the 248 clips above.

\paragraph{SongEval.} Each judge hears 30~s of each song, from 30~s into the song (the first 30~s if the song is shorter than 60~s), mono at 16~kHz. The rating instruction (first wording) is: ``You will hear a song produced by an AI music generator. Rate the overall musical quality of the song on a scale from 1 (very poor) to 5 (excellent). Answer with a single digit.'' The second and third wordings are ``The audio is a piece of music produced by an AI music generator. How good is the music overall? Give a score from 1 to 5, where 1 is very poor and 5 is excellent. Reply with one digit only.'' and ``Act as an expert music critic and listen to the generated song. Score its overall musical impression from 1 (worst) to 5 (best). Output only the score.'' The pairwise instruction is: ``Two songs were produced by AI music generators. You will hear them one after the other, separated by a short beep: song A first, then song B. Which song has the better overall musical quality? Answer with a single letter, A or B.'' The 300 order-swap pairs are drawn at random from all songs with a fixed seed (0) and cover 525 songs; each song enters as the first 12~s of its excerpt, and the two excerpts are joined by a 0.6~s tone.

\end{document}